\pdfoutput=1

\documentclass[11pt]{article}

\usepackage{acl}

\usepackage{times}
\usepackage{latexsym}
\usepackage{tikz}
\usetikzlibrary{shapes.geometric, arrows}

\usepackage[T1]{fontenc}

\usepackage[utf8]{inputenc}

\usepackage{microtype}

\title{Should I disclose my dataset?\\Caveats between reproducibility and individual data rights}

\author{Raysa M. Benatti \\
  Institute of Computing \\
  University of Campinas \\
  Campinas, Brazil\\
  \texttt{\small raysa.benatti@gmail.com} \\\And
  Camila M. L. Villarroel \\
  Law School of Ribeirão Preto\\
  University of São Paulo\\
  Ribeirão Preto, Brazil\\
  \texttt{\small cami.lima.v@gmail.com} \\\AND
  Sandra Avila \\
  Institute of Computing \\
  University of Campinas  \\
  Campinas, Brazil\\
  \texttt{\small sandra@ic.unicamp.br} \\\And
  Esther L. Colombini \\
  Institute of Computing \\
  University of Campinas  \\
  Campinas, Brazil\\
  \texttt{\small esther@ic.unicamp.br} \\\And
  Fabiana C. Severi \\
  Law School of Ribeirão Preto\\
  University of São Paulo\\
  Ribeirão Preto, Brazil\\
  \texttt{\small fabianaseveri@usp.br}}

\begin{document}
\maketitle
\begin{abstract}
Natural language processing techniques have helped domain experts solve legal problems. Digital availability of court documents increases possibilities for researchers, who can access them as a source for building datasets --- whose disclosure is aligned with good reproducibility practices in computational research. Large and digitized court systems, such as the Brazilian one, are prone to be explored in that sense. However, personal data protection laws impose restrictions on data exposure and state principles about which researchers should be mindful. Special caution must be taken in cases with human rights violations, such as gender discrimination, over which we elaborate as an example of interest. We present legal and ethical considerations on the issue, as well as guidelines for researchers dealing with this kind of data and deciding whether to disclose it.
\end{abstract}

\section{Introduction}
\label{intro}

The increasing availability of data in digital formats, along with the means to process and interpret it, has boosted the interest in its versatile use. The enriched commercial value of personal data has justified the adoption of personal data protection laws aiming to protect individual and collective rights. Having such a legal structure --- with a broader social recognition of implications associated with personal data usage --- demands that data controllers be mindful about ethical issues and legal liabilities when dealing with this resource.

Research agents have been major controllers of data on individuals. While science has always relied on data, the societal switch to digital-intensive structures has changed much of their nature, amount, and availability. This context calls for specific approaches from researchers when balancing individual rights and scientific reproducibility --- since disclosing datasets, while beneficial for research publicity, might expose information over which special considerations might apply. 

Computational research based on data-intensive frameworks, such as machine learning, typically operates over collecting, processing, and interpreting large amounts of data; being used to awareness of resource sharing, the computing community tends to encourage reproducibility practices. In experimental contexts, that usually means disclosing descriptions of methods and results and codes, tools, and data. 

Digital data can come from many sources. When derived from the realm of the social sciences it is often produced in text form, which motivates its use as input for natural language processing methods. Social scientists have relied on computational approaches to help answer some of their research questions; in the legal domain, court documents often provide rich material, which computational tools allow to be analyzed on improved scales.  

Among the many inquiries that large-scale analysis of court documents could help address, we are particularly interested in gender-related ones. Examples include: (a) Which role do gender biases play in decisions regarding gender-based violence (GBV) legal cases? (b) How many cases are linked to the same victim? (c) How many police investigations make it to court? These are research questions for which natural language processing methods seem suitable.  

Domain experts often identify demands for this research while exploring their own areas, creating communities around common issues. A large community of researchers and practitioners interested in how computational approaches can be used to address questions in the legal domain has emerged in Brazil; the country is one of the most litigious of the world in the court, having one lawyer for each batch of around 160 people\footnote{\href{https://www.oab.org.br/institucionalconselhofederal/quadroadvogados}{Data} from the Brazilian Bar Association (\textit{Ordem dos Advogados do Brasil}).}, and approximately 80 million active legal cases\footnote{Data from the \href{https://anuario.conjur.com.br/pt-BR/profiles/78592e4622f1-anuario-da-justica/editions/brazil-justice-yearbook-2022}{2022 Brazil Justice Yearbook}.}.

With a substantial court system, large databases of documents issued by such courts, and an engaged research community in the field, Brazil emerges as a legal data hotspot --- with many issues regarding data disclosure from state entities and researchers. The country issued its General Data Protection Act in 2018, based on European's General Data Protection Regulation, which expanded the debate on such issues. 

Focus on GBV-related cases is justified not only by research and human rights significance but also due to the amount of delicate personal information they carry on the subjects involved, meaning that disclosing them without regard for legal and ethical principles could implicate severe harm. While focusing on this context, we stress that our considerations might apply to others.

A similar observation should be made for the location we chose to highlight. Focusing on the Brazilian context will benefit its large community of researchers and practitioners interested in the field. It may also provide useful insights from other legal settings --- particularly civil law ones (e.g., continental Europe), in which Brazilian legal structures and fundamental statutes are heavily based.

Our main contributions are:
\begin{enumerate}
    \item To bring ethical considerations on personal data disclosure by researchers;\vspace{-0.1cm}
    \item To provide guidelines for researchers to help them decide on data disclosure;\vspace{-0.1cm}
    \item To discuss how to preserve both reproducibilities of computational research and individual data rights.
\end{enumerate}
We hope to help the community of interested researchers and practitioners understand the fundamentals of the Brazilian data protection legal system and its caveats.

This paper is organized as follows. Section \ref{repro} introduces research reproducibility concepts. It is followed by discussions on data disclosure and publicity in Section \ref{legal}, where we present legal principles, practical issues, and ethical concerns on the matter. Risk assessment and mitigation measures are described in Section \ref{risk}. Sections \ref{legal} and \ref{risk} also suggest guidelines of good practices for researchers. Finally, Section \ref{path} summarizes approaches that could help researchers address concerns on disclosure of court documents and similar data. 

\section{Reproducibility}
\label{repro}

Reproducibility has been at the core of the debate on scientific integrity, being recognized as a critical quality of modern research \cite{goodman-2016, baker-2016, loscalzo-2012}. The concept is open enough to evoke debate on its meaning, and comprises different aspects of scientific soundness and accountability \cite{goodman-2016, drummond-2009}; however, there appears to be some consensus on the importance of community scrutiny for research quality assessment, for which reproducibility is essential. 

Scrutiny, fraud prevention, and fraud detection are not the only motivation behind efforts toward reproducible research. Science is a collective endeavor of public interest; therefore, resource-sharing strengthens networks, creates research possibilities, and helps build connections inside and between communities --- not only for science itself but also for practitioners and society as a whole.   

This is especially true in empirical research, as is usually the case in computer science. In fact, many efforts have been made towards fostering a culture of openness of resources inside the computing-related community --- from free and open source software initiatives\footnote{\url{https://www.fsf.org}}\footnote{\url{https://opensource.org}} to open science guidelines and frameworks \cite{wilkinson-2016, peng-2011, sonnenburg-2007}. \citet{peng-2011} describes a reproducibility spectrum for computer science research in which the gold standard would be attained by publishing linked and executable code and data along with results. In some fields, such as machine learning, the importance of empirical choices behind results that might support decision-making processes is such that it could justify one arguing that reproducibility is as important of a property as the research results themselves.

In this context, data sharing and quality assessment emerge as an object of concern as well \cite{gebru-2021, deschutter-2010, blockeel-2007}. Data collecting, cleaning, labeling, and/or processing are often part of the experimental pipeline in machine learning research, which justifies interest in making them available for peers and stakeholders. In some cases, however, the means and extent to which data should be shared are not trivial decisions. 

When individual rights of the subjects regarded in the dataset might be at stake, sharing this data becomes a challenge since adjustments --- or even the decision not to share --- might be needed to avoid legal and/or ethical violations. Privacy, for instance, is one of the main concerns \cite{proell-2015}. Some domains, such as health and clinic research, are notably prone to this issue. When faced with such a situation, researchers must take legal and ethical boundaries into account, assess the risks involved in disclosing the data, and weigh them against the benefits of reproducibility.

\section{Issues on disclosing data provided by courts}
\label{legal}

Particularities on data sharing emerge in the context of research that uses computational approaches to court decisions. This section delves into some of them from the perspective of our research example: exploring natural language processing and other computational techniques over Brazilian court decisions in GBV-related cases. However, as mentioned in Section \ref{intro}, our considerations might also be helpful for other contexts. 

\subsection{Publicity vs. Reproducibility}
\label{publicity}

Brazilian court decisions are, by default, public documents. Publicity\footnote{Meaning, in this context, transparency or openness.} of procedural acts issued by the justice system is such an important principle that it is stated in the country's federal constitution (articles 5 LX, 93 IX, and 93 X), which provides secrecy as an exception to be reserved for the protection of ``intimacy'' and ``social interest''. (Secrecy is discussed further in Section \ref{secrecy}.) Codes of civil (articles 11 and 189) and criminal (article 792) procedures, which present bounding proceeding rules for legal cases, have similar statements. 

The National Council of Justice (CNJ)\footnote{\url{https://www.cnj.jus.br}}, created in 2004 to supervise and manage the Brazilian justice system, provides more specific regulations on the matter. It declares that essential data regarding legal cases must be publicly accessible to ``any person, regardless of previous enrollment or demonstration of interest'' (Res. 121, article 1). The list of what is considered to be essential data includes (a) number, class, and theme; (b) name of parties and their lawyers; (c) procedural flow and updates; (d) full content of court decisions. Other documents, such as petitions and investigation reports, are restricted to lawyers, parties, and some official entities (articles 2 and 3). Again, cases that must remain in secrecy are preserved as exceptions. 

Some provisions foster the use of digital documents in the justice system rather than physical ones, such as Federal Law 11419/2006 and regulation from the CNJ itself (Res. 215, articles 5 and 6). This scenario increases the availability of data for computational research purposes since it facilitates the extraction and processing of legal information. In the context of our research example and similar ones, it is then possible to scrape such documents and build datasets based on them --- along with metadata, executable code, and research results, attaining a gold standard of scientific reproducibility. In that sense, we could acknowledge reproducibility as analogous to publicity, perceiving reproducibility as the public sector publicity principle applied to the science realm. Ultimately, they are both cultivated in the name of the public interest behind their related activities, which requires scrutiny, transparency, and community implication in their processes.

However, we recognize caveats. It does not follow from court decisions being publicly available by default that researchers could relinquish concerns when scraping and building datasets from these documents; our research example can illustrate that, as described in Sections \ref{secrecy}, \ref{restrictions} and \ref{risk}.

Despite the intersection between motivations supporting publicity and reproducibility, the justice system has different obligations and prerogatives than research institutions. When disclosing a court decision, the state complies with a legal duty to publicize and acts by itself; it claims the rights and responsibilities carried by such a publicization. If another person or entity --- for instance, a researcher or research agency --- extracts and discloses the same record, s/he creates another point of access, claiming responsibility over the content (even if unwittingly).

Another issue arises in that, in research settings, the data might not be shared on its own; instead, it is often made available in the context of an experimental pipeline, with annotation, modifications, associated code, and/or results from models learned from them. In that case, disclosing the data is more than merely indexing it; it also publicizes it from a specific perspective. It makes sense that whoever is in charge of disclosing it is also legally and ethically responsible. Thus, when seeking reproducibility, researchers must account for that boundaries, being wary about emulating publicity-guided acts from the public administration.   

\subsection{The issue(s) of secrecy}
\label{secrecy}

Access to information is a fundamental right in a democratic environment. In Brazil, its legal and constitutional strengthening is linked to democratization processes in the 80s and later, after the country's military dictatorship. The right to information is a fundamental element of civic citizenship and scrutiny of executive, legislative, and judiciary spheres of power, protected by several national and international legal statements.

In addition to the default public status of court decisions, transparency propositions also apply to documents provided by public institutions in general (LAI\footnote{Legal abbreviations are described in \ref{sec:appendix} (Appendix).}, articles 2 and 3), and publicity is a vital principle of public administration (CF, article 37). Therefore, confidentiality\footnote{Although secrecy and confidentiality have the same meaning, we can interpret secrecy (a concept mainly used in the context of the justice system) as a type of confidentiality (that can apply to any document, data, or information).} is an exception and must be justified by legal restrictions and/or particular circumstances --- such as when national security is at risk (LAI, articles 3 III and 23).

In some cases, publicity and open access to information are restricted due to the need to protect other important rights or principles --- notably intimacy and social interest (CF, article 5 LX). Intimacy, personal life, honor, and image are individual rights protected by the federal constitution (article 5 X) and other statements, such as the Access to Information Act (article 31). However, confidentiality must be well justified due to the (theoretically) quasi-paramount status of publicity-based principles in the Brazilian legal system. 

\paragraph{When is secrecy justified?} In Brazilian civil cases, the law states specific circumstances that warrant secrecy: (a) if needed to preserve matters of social or public interest; (b) in disputes on marriage, separation, divorce, civil union, parentage, alimony, or custody of children and adolescents; (c)~in cases with data protected by the constitutional right to intimacy; (d) in arbitration cases (CPC, article 189). Interpretation of these statements is usually restrictive for the benefit of publicity.  

In the criminal realm, secrecy is legally established in all crimes against sexual dignity (CP, article 234-B). The judge might also declare secrecy on a criminal case to avoid the victim's exposure to the media (CPP, article 201, 6\textsuperscript{th} paragraph). 

Besides legal restrictions, any party of a dispute has the right to request secrecy on the whole case or on specific documents, which might or not be granted by the judge --- who also has the authority to revoke it, \textit{ex officio} or by request (CNJ Res. 185, article 28). 

This set of rules means that secrecy is established in many GBV-related lawsuits, since family law, civil disputes, and cases on sexual crimes are notably settings where gender-based abuse and biases are often brought to court. Therefore, when dealing with court decisions in this domain, one must be attentive to confidentiality boundaries that might restrain data disclosure.  

\paragraph{Who can access these court decisions?} When secrecy is established, court documents --- including usually public ones such as decisions --- are only accessible to parties and their lawyers (CNJ Res. 121, article 1)\footnote{It is granted that they are also available to the justice system employees whose work is operationally essential for the case to be processed, e.g., the assigned judge.}. Secrecy is also a legal exception to the general rule of access to information (CNJ Res. 215, article 12 VII; LAI, article 22). 

Courts might establish internal rules to deal with different degrees of secrecy --- e.g., some cases might be totally unavailable except for allowed people, while others might have some documents publicized as long as information on parties is previously anonymized. However, such anonymization does not always happen as expected, especially in large courts where the systematization of documents is particularly challenging. In that case, decisions that are supposed to remain in total secrecy can end up publicly available. While courts are liable for the publicization, and it is not reasonable to expect researchers always to identify when that is the case, they should be aware of this possibility. 

\paragraph{Guidelines of good practices} Given the restrictions derived from secrecy in some legal cases, researchers might consider the following guidelines of good practices for data disclosure when working with datasets made of court documents:

\begin{itemize}
    \item If data is provided from secrecy cases, it \textbf{should not} be disclosed \textbf{unless} it is thoroughly anonymized and/or provided by demand only, with a deed of undertaking (details in Section \ref{risk});\vspace{-0.1cm} 
    \item Otherwise, the researcher should check if other restrictions apply (Section \ref{restrictions}). 
\end{itemize}

We stress that having been able to access court decisions online does not guarantee that the case is not under secrecy. Deciding to disclose non-anonymized secret documents is a legal liability since it might violate privacy and intimacy rights, subjecting the liable person or entity to penalties. 

\subsection{Personal data restrictions}
\label{restrictions}

Court documents might carry publishing restrictions justified by reasons beyond secrecy, especially since personal data of people involved in legal cases are often disclosed in this material. Recent data protection laws, such as Brazil's General Data Protection Act (LGPD) and Europe's General Data Protection Regulation (GDPR), emerged in the context of increasing commercial usage of (more abundant than ever) personal data; thus, their main goal is to protect individuals from potentially abusive behavior perpetrated by profit-oriented agents. Legal restrictions on personal data usage are not the same for agents who do not fall under this category, such as public institutions and researchers; however, liabilities and ethical issues might still apply to them.

\begin{figure}[t]
    \centering
    \includegraphics[width=0.35\textwidth]{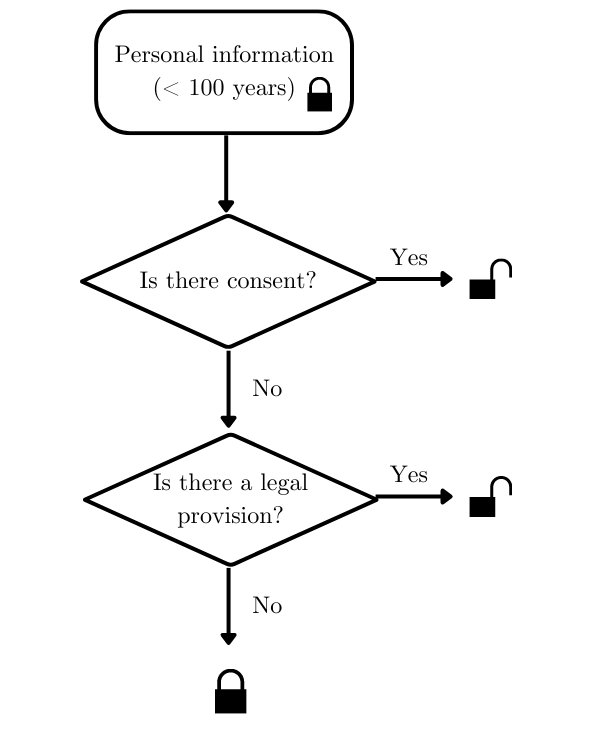}
    \caption{Flowchart of incidence of Access to Information Act restrictions (article 31) on personal information.}
    \label{fig:fc1}
\end{figure}

In Brazil, the concept of personal information precedes LGPD; the Access to Information Act defines it as ``information regarding identified or identifiable natural person'' (article 4 IV) and states restrictions on its processing\footnote{\label{tratamento}Processing (\textit{tratamento}) refers to ``any operation or set of operations which are performed on personal data or sets of personal data, whether or not by automated means'' (GDPR, article 4(2)). It can mean use, storage, diffusion, destruction, alteration, collection, retrieval, extraction, and so forth. Thus, it might include any operation in a machine learning pipeline --- collecting, cleaning, using as input for models, publishing.} (article~31). Figure~\ref{fig:fc1} shows a flowchart on whether personal information can be processed (open padlock); it applies to personal information whose production happened not earlier than 100 years ago --- since, in that case, confidentiality no longer applies\footnote{Lifting confidentiality after a maximum of 100 years allows for the use and interpretation of documents regarding their historical value since cultural heritage is a protected asset under the federal constitution (article 216).} (article~31, 1\textsuperscript{st}~paragraph, I). 

Personal information can be processed \textbf{if} there is explicit consent from the owner of its rights \textbf{or} if there is a legal provision to do so. In computational research settings, getting consent from all subjects involved is seldom feasible; therefore, if willing to abide by this statute, researchers might consider if their use case can be framed as a legally supported~exception. 

Usually, it can. The Act presents statistical and scientific research of ``evident public or general interest'' as a situation allowing personal information processing without the need for consent --- as long as anonymization is guaranteed. Other exceptions include: (a) for medical treatment if the owner of rights is incapable of consenting; (b) to fulfill a court order; (c) if necessary for the defense of human rights; (d) to protect the public and general interest. We argue that scientific activity itself is a matter of public interest; therefore, not only could it be framed in hypothesis (d) (which would dismiss the need for data anonymization), it is redundant to require evidence of public interest to allow for information processing in this case. 

\begin{figure}[t]
    \centering
    \includegraphics[width=0.58\textwidth,trim={1.25cm 2cm 0 0},clip]{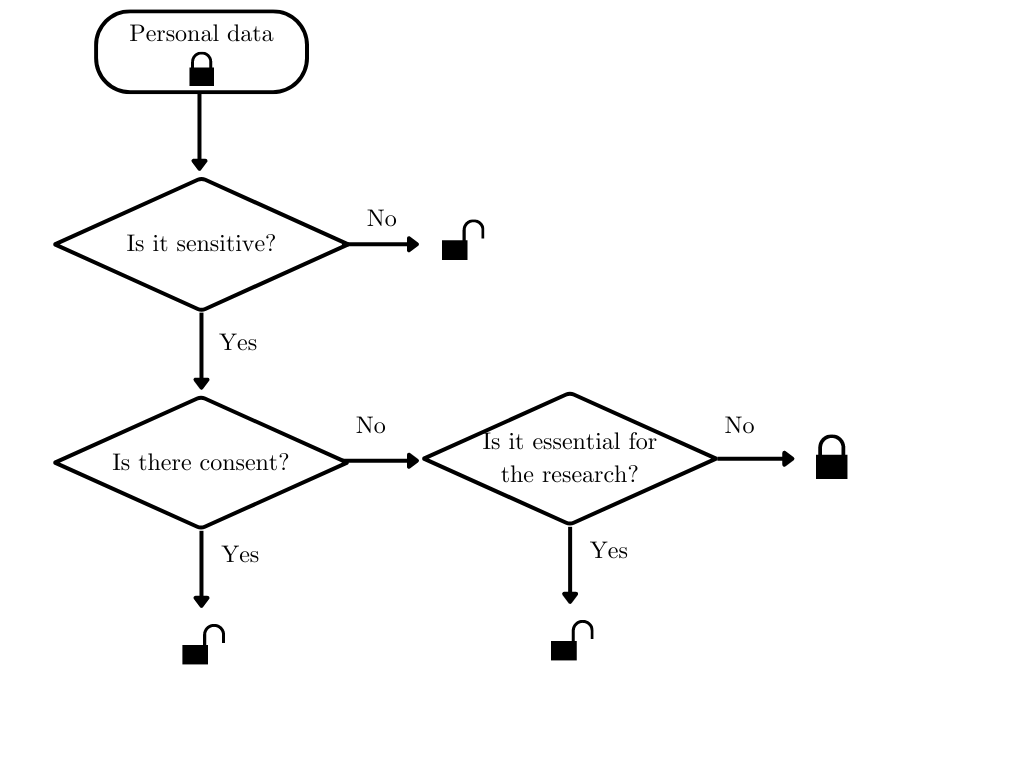}
    \caption{Flowchart of incidence of LGPD restrictions for researchers (articles 7 and 11) on personal data processing.}
    \label{fig:fc2}
\end{figure}

In our study scenario, demanding anonymization also conflicts with what is stated by the LGPD --- according to which it would be optional, although recommended. Figure \ref{fig:fc2} shows a flowchart for researchers willing to comply with this statute regarding processing personal data. Research settings entail a special application of the law (article 4 II (b)), being one of the situations in which personal data might be processed (article 7 IV) and conserved (article 16 II) as long as: 

\begin{itemize}
    \item Data is \textbf{not} sensitive \textbf{and} general principles of the law, as well as function, good faith and public interest, are preserved; \textbf{or}\vspace{-0.1cm} 
    \item There is consent from the owner of rights; \textbf{or}\vspace{-0.1cm} 
    \item The operation is essential for the research activity to be performed. 
\end{itemize}

In any case, anonymization must be assured ``whenever possible''. Thus, it is not a duty, but a recommendation, not entailing punishment if not followed --- which means that complying with it is an ethical deed of the researcher rather than a legal~obligation. 

Personal data is sensitive if it refers to racial or ethnic origin, political opinions, religious or philosophical beliefs, trade union membership, health or sex life, or personal genetic or biometric information --- as stated in article 9(1) of GDPR, with a similar provision in Brazilian law. Sensitiveness of data implies special responsibilities for its processing; for researchers, processing of sensitive data can only occur if (a) there is consent from the owner of the rights or (b) the operation is \textbf{essential} for the activity. 

Once a research project has been designed, and the need for using sensitive data in its context has been demonstrated, indispensability is established --- therefore complying with legal provisions. There remains, however, the issue of whether full reproducibility is an imperative element of the scientific endeavor that would justify disclosing sensitive data to fulfill essential research activities. We argue that preserving sensitive data, while might diminish possibilities of replicability, does not hamper acceptable levels of reproducibility \cite{drummond-2009}; thus, when using this data, disclosing it only under mitigation guidelines (as described in Section \ref{risk}) might be a fair trade-off between research publicity and protection of human rights. While the most usual metadata provided with court decisions (e.g., names of parties and their lawyers) are not sensitive, such documents might contain information that, when combined with identification of parties, becomes sensitive --- even when not issued in cases under secrecy. This arises since court sentences must include a report on the case and the reasoning behind the verdict\footnote{These are required elements for any court sentence issued under Brazilian law, besides the verdict itself (CPC, article 489); other legal systems have similar provisions \cite{neto-2022}.} --- which could contain sensitive information on the subjects\footnote{As an example, if a domestic violence case is brought to court and issues on the sex life of the people involved are relevant for the circumstances, such issues will possibly be described in the decision report and/or motivation --- thus exposing sensitive information on the identifiable subjects.}. 


In other situations, while the legal case is not under secrecy nor displays strictly sensitive information, other forms of delicate information might appear in a court decision. For instance, in domestic violence cases, children and/or teenagers often witness the event and are either listened to in court or mentioned in case reports, therefore having their names (or other data) exposed in public documents. While there might not be an explicit legal restriction for researchers to fully disclose such records, doing so would raise ethical concerns. 

\section{Risk assessment and mitigation}
\label{risk}

When faced with the decision to disclose court documents used in research, one must confront risks against the benefits of science replicability since full disclosure might potentially harm and violate the rights of the subjects whose personal data is displayed. Risks can exist regardless of legal restrictions, given that records from courts typically carry a large amount of personal information of parties, witnesses, and other subjects related to the case, both in the document(s) text and metadata. 

Making personal data available establishes as liable the person or entity in charge of the disclosure, who becomes a controller according to law (GDPR, article 4(7); LGPD, article 5 VI) \cite{lenora-talk}. As a controller, a researcher or research agency operates under distinct ethical guidelines than those of courts and law enforcement agencies --- which, when disclosing personal data, are usually complying with a legal duty to transparency and publicity, as well as broader public interest principles. While carrying public interest on its own, science reproducibility is not a legal obligation (thus not dismissing liability in the same way that applies to state entities), and can be acceptably achieved with mitigation-mindful data availability when full disclosure is not allowed or advised.

Further, the legal system context represents a special circumstance for personal data disclosing due to implications regarding rights of access to justice, due process of law, and defense --- which also relates to publicity and transparency. One would be unable to build a defense if not provided with complete information on the case, including data on parties and their lawyers, allegations, documents, and evidence. Transparency of court documents is generally a matter of state accountability. Imposing severe constraints against this kind of publicity could have noxious outcomes for democratic settings and is not the same as restricting personal data disclosure in scientific frameworks. 

In that sense, although documents used in research might be publicly available in other sources (e.g., court websites), their disclosure by researchers can increase risks for the subjects involved, considering that: (a) it reunites the data in a single, cohesive source, often cleaner, and more structured than the original and combined with annotation and metadata, therefore making it easier for different groups of people to access it and make inferences from it; (b) public status of such documents in original sources might change over time, adding an extra layer of harm-related responsibility on the researcher who decides to disclose them.

In the context of GBV-related cases, risks of full personal data disclosure by researchers or research agencies include:

\begin{itemize}
    \item Violation of privacy and intimacy rights of:\vspace{-0.1cm}
    \begin{itemize}
        \item minors, in disagreement with their best interest and right to informational self-determination;\vspace{-0.1cm} 
        \item victims and witnesses, which might contribute to reinforcing their vulnerability against aggressors and their communities;\vspace{-0.1cm}
        \item defendants, which might contribute to reinforcing penal populism actions and ideas at the cost of individual rights violations;\vspace{-0.1cm}
    \end{itemize}
    \item Exposure of sensitive data, which might violate the civil rights of the subject(s);\vspace{-0.1cm}
    \item Exposure of confidential information;\vspace{-0.1cm}
    \item Exposure of any information that might jeopardize the safety or integrity of the subject(s) involved in a legal case.  
\end{itemize}

In fact, such risks have been used to advocate for initiatives such as Bill 3333/20, whose main proposal is to establish ``absolute secrecy'' for personal information displayed in police reports and court documents in cases of domestic violence --- which are currently public by default. If approved, alleged aggressors would be hindered from accessing personal data on the victim(s), thus impairing their right to defense. For researchers, this would add a class of documents in the secrecy-justified caution cluster. 

Exposing sensitive and/or confidential data can increase the possibilities of rights restrictions, retaliation from a subject's community and institutions, and physical and mental suffering. Let us consider, for instance, the disclosure of the LGBTQ+ status of a subject implicated in a legal case: such a deed could have discrimination-related consequences such as the loss of a job, impairment of social and family ties, or threats to one's physical integrity.

Ease of access to data obtained from courts allows for inferences that would hardly be made otherwise --- an exciting possibility for good-faith researchers and policymakers but also a caution-inspiring scenario. From a dataset of Brazilian court decisions with specific characteristics, for example, one could extract a map of precise addresses of victims, defendants, or plaintiffs (some of which could be minors or belong to other protected groups). An ill-motivated, technically capable agent could use that information to perpetrate physical, moral, emotional, or other kinds of harm to these people --- and, while there are legal provisions to make perpetrators accountable, some damages might be beyond repair. 

We note that the risks mentioned above do not constitute an exhaustive list; ideally, researchers should evaluate which issues might apply to their context and know their data enough to build a proper risk assessment in order to decide on the extent of data disclosure considering available resources and both ethical and legal restrictions. 

When personal information is part of the data source in research, mitigating such risks is possible and advised. Risks are usually associated with data disclosure rather than their use itself. Personal data protection laws ordinarily do not distinguish use from disclosure for legal purposes, placing both operations under the concept of ``processing'' (see Note \ref{tratamento}). However, discerning them is relevant in our context of interest. 

While using court documents in research settings (e.g., as input for training models or to perform other quantitative and qualitative analyses) does not directly threaten or pose harm to subjects involved, disclosing them without taking prior mitigation actions might do. We identify three levels of personal data implication for our context:

\begin{enumerate}
    \item \textbf{From secret cases}: Not to be disclosed without mitigation; disclosure without mitigation both legally and ethically inappropriate;\vspace{-0.1cm}
    \item \textbf{From non-secret cases, with sensitive data}: Not illegal for researchers to disclose without mitigation if the disclosure is essential for research; disclosure without mitigation might be ethically debatable;\vspace{-0.1cm}
    \item \textbf{From non-secret cases, without sensitive data}: Not illegal for researchers to disclose; disclosure without mitigation should ideally be preceded by an analysis of specific context and risk-benefit assessment. 
\end{enumerate}

Mitigation measures to protect personal data embedded in public court documents might include several actions from researchers and research agencies, who should evaluate the risks of data disclosure, benefits of full replicability, and availability of resources to perform mitigation. We stress two of them: (a) anonymization and (b) disclosure by demand with a deed of the undertaking. 

\paragraph{Anonymization} When personal data is anonymized, it is no longer considered personal data (LGPD, article 12; GDPR, recital 26) --- therefore, none of the issues discussed in this work would apply, and documents containing them could be disclosed, \textit{ab initio}, without legal nor ethical implications. To be considered fully anonymized, personal identification corresponding to the data must be untraceable and not reversible by reasonable efforts\footnote{What could be considered ``reasonable'' is open for debate and can vary depending on specifics of each case, as explained by Vokinger et al. \citeyearpar{vokinger-2020}.}; thus, pseudonymization --- which allows for identification to be restored ---, while allowed to comply with legal guidelines on data storage, is not enough to allow full disclosure. 

There are, however, practical obstacles. Full anonymization is not always attainable since it might require massive manual efforts or the use of technically challenging tools, which do not necessarily guarantee complete accuracy. Some kinds of data are challenging to anonymize; computational research often deals with large amounts of documents and sensitive information is usually non-structurally embedded in the text, meaning that masking them pre-disclosure --- or even identifying them --- might not be possible. Deeper discussions on technical and juridical aspects of legal data anonymization can be found in the works of Csányi et al. \citeyearpar{gergely-2021} and van Opijnen et al. \citeyearpar{opijnen-2017}.

Regarding replicability, anonymization barely affects it unless the personal information is relevant for the analysis. In some cases, determining the relevance of personal information for experimental settings is overly demanding and/or outside of the scope of research, e.g., when black-box models learn from input documents. In those scenarios, approaches for model interpretability and/or explainability might be taken into consideration \cite{rudin-talk, rudin-2019, molnar-2022}. At any rate, if research results and code are duly published and the methodology is thoroughly explained, reproducibility should not be severely disturbed. Assuming that the documents used as the source are publicly available, anyone following the same procedures should be able to access them, therefore claiming their responsibility upon processing the data. 

If mitigation is needed or advised, but adequate anonymization is not feasible, researchers should consider mitigation measures described next.

\paragraph{Disclosure by demand} In this case, the person, group, or entity responsible for research provides a contact channel through which the data can be requested and sent by demand. Ideally, whoever requests the material should agree to a deed of undertaking bound by the good faith of parties, with clauses preventing inappropriate data processing and protecting the subjects' best interest. Traceability of data controllers is a major advantage of this~method.

While being the safest option regarding personal data protection, we identify the following caveats: (a) it relies on assuming good faith of the researchers; (b) it constrains replicability, given that it adds extra layers of compromise, bureaucracy, and communication for interested parties. 

Also, mitigation measures (a) and (b) could be combined, although this would require extra effort. Researchers can still decide not to make data available, therefore escaping from the burden of responsibility over the dataset disclosure and choosing privateness over publicity.  

\section{Possible paths}
\label{path}

Both research reproducibility and data protection of subjects mentioned are essential values in democratic settings and must be preserved and encouraged. Good research practices and awareness of legal and ethical restrictions can help researchers and agencies decide whether --- and to which extent --- disclose their court documents datasets. While much of the responsibility for the form and availability of such documents relies on the courts, researchers also have liability over the content they choose to disclose. The following approaches could help them address it in the future. 

\begin{description}
    \item [Guidelines:] While provisions for researchers should not be too strict, having more explicit guidelines or recommendations in place --- provided by national authorities on data protection and other official entities --- could help address some of the concerns;
    \item [Anonymization tools:] Adequate anonymization of data is not trivial. While this burden does not rely solely on researchers, tools that help get past this task might encourage them to act in this sense;
    \item [Official data repositories:] Much of current replicability practices rely on individual data repositories. Having official, institutional data repositories in place, backed up by research agencies and supplemented by somewhat automatic deeds of undertaking by parties, could be an option for data availability without compromising protection of individual data rights. 
\end{description}



We expect that, with proper guidelines of good practices and tools, as well as engagement from the scientific community and state agencies, a fair balance can be achieved between the publicity that guides research and the protection of human rights~and the informational self-determination of~individuals. 





\section*{Acknowledgements}


R.~Benatti is partially funded by CAPES/Brazil and FAEPEX/Unicamp. 
C. Villarroel is partially funded by FAPESP (São Paulo Research Foundation) 2017/26174-6.
F. Severi is supported by the University of São Paulo's Law School of Ribeirão Preto. 
S. Avila is partially funded by CNPq/Brazil 315231/2020-3, FAPESP 2013/08293-7, 2020/09838-0, H.IAAC (Artificial Intelligence and Cognitive Architectures Hub), and Google LARA 2021. E. Colombini is partially funded by CNPq PQ-2 grant 315468/2021-1 and H.IAAC. 
The Recod.ai lab is supported by projects from FAPESP, CNPq, and CAPES.
Finally, we are grateful for the reviewers of this work, who enriched it with their evaluation and comments.

\bibliography{anthology,custom}

\appendix

\section{Appendix: List of legal statutes mentioned in this paper}
\label{sec:appendix}

In order of appearance:

\begin{enumerate}
    \item \href{http://www.planalto.gov.br/ccivil_03/constituicao/constituicaocompilado.htm}{CF} (\textit{Constituição Federal}): Brazilian Federal Constitution (1988);\vspace{-0.1cm}
    \item \href{http://www.planalto.gov.br/ccivil_03/_ato2015-2018/2015/lei/l13105.htm}{CPC} (\textit{Código de Processo Civil}): Brazilian Code of Civil Procedures (Law n. 13105, March 16, 2015);\vspace{-0.1cm}
    \item \href{http://www.planalto.gov.br/ccivil_03/decreto-lei/del3689compilado.htm}{CPP} (\textit{Código de Processo Penal}): Brazilian Code of Criminal Procedures (Decree-Law n. 3689, October 3, 1941);\vspace{-0.1cm}
    \item \href{https://atos.cnj.jus.br/atos/detalhar/atos-normativos?documento=92}{CNJ Res. 121}: National Council of Justice, Resolution n. 121 (October 5, 2010);\vspace{-0.1cm}
    \item Brazilian \href{http://www.planalto.gov.br/ccivil_03/_ato2004-2006/2006/lei/l11419.htm}{Law n. 11419/2006} (December 19, 2006);\vspace{-0.1cm}
    \item \href{https://atos.cnj.jus.br/atos/detalhar/2236}{CNJ Res. 215}: National Council of Justice, Resolution n. 215 (December 16, 2015);\vspace{-0.1cm}
    \item \href{http://www.planalto.gov.br/ccivil_03/_ato2011-2014/2011/lei/l12527.htm}{LAI} (\textit{Lei de Acesso à Informação}): Brazilian Access to Information Act (Law n. 12527, November 18, 2011);\vspace{-0.1cm}
    \item \href{http://www.planalto.gov.br/ccivil_03/decreto-lei/del2848compilado.htm}{CP} (\textit{Código Penal}): Brazilian Criminal Code (Decree-Law n. 2848, December 7, 1940);\vspace{-0.1cm}
    \item \href{https://atos.cnj.jus.br/atos/detalhar/1933}{CNJ Res. 185}: National Council of Justice, Resolution n. 185 (December 18, 2013);\vspace{-0.1cm}
    \item \href{http://www.planalto.gov.br/ccivil_03/_ato2015-2018/2018/lei/l13709.htm}{LGPD} (\textit{Lei Geral de Proteção de Dados}): Brazilian General Data Protection Act (Law n. 13709, August 14, 2018) -- also available \href{https://iapp.org/media/pdf/resource_center/Brazilian_General_Data_Protection_Law.pdf}{in English} (unofficial translation);\vspace{-0.1cm}
    \item \href{http://data.europa.eu/eli/reg/2016/679/2016-05-04}{GDPR}: European General Data Protection Regulation (Regulation (EU) 2016/679 of the European Parliament and of the Council of 27 April 2016);\vspace{-0.1cm}
    \item \href{https://www.camara.leg.br/propostas-legislativas/2255231}{Bill 3333/20}: Brazilian Chamber of Deputies, Bill (\textit{Projeto de Lei}) n. 3333 (2020); author: deputy Ricardo José Magalhães Barros. 
\end{enumerate}

\end{document}